\documentclass[12pt]{article}
\usepackage[british]{babel}
\usepackage{epsfig}

\begin{document}                                                                
\date{}

\title{
{\vspace{-25mm} \normalsize
\hfill \parbox{50mm}{DESY 02-146}}\\[10mm]
On the price of light quarks
\footnote{Contribution by I. Montvay to QUARKS'2002,
          Novgorod, Russia, June 2002} \\[1em] }

\author{qq+q Collaboration \\[0.5em]
        F. Farchioni$^{\scriptscriptstyle a}$, 
        C. Gebert$^{\scriptscriptstyle b}$,
        I. Montvay$^{\scriptscriptstyle b}$,
        L. Scorzato$^{\scriptscriptstyle b}$}

\newcommand{\be}{\begin{equation}}                                              
\newcommand{\ee}{\end{equation}}                                                
\newcommand{\half}{\frac{1}{2}}                                                 
\newcommand{\rar}{\rightarrow}                                                  
\newcommand{\lar}{\leftarrow}
\newcommand{\LCB}{\raisebox{-0.3ex}{\mbox{\LARGE$\left\{\right.$}}}
\newcommand{\RCB}{\raisebox{-0.3ex}{\mbox{\LARGE$\left.\right\}$}}}
                                                                                
\maketitle
\vspace*{-2em}
\begin{center}{\large
$^{\scriptstyle a}$
Westf\"alische Wilhelms-Universit\"at M\"unster, \\
Institut f\"ur Theoretische Physik,\\
Wilhelm-Klemm-Strasse 9, D-48149 M\"unster, Germany \\[1em]
$^{\scriptstyle b}$
Deutsches Elektronen-Synchrotron DESY    \\
Notkestr.\,85, D-22603 Hamburg, Germany}
\end{center}
\vspace*{1em}

\begin{abstract} \normalsize
 The computational cost of numerical simulations of QCD with light
 dynamical Wilson-quarks is estimated.
 The qualitative behaviour of the pion mass and coupling at small
 quark masses is discussed.
\end{abstract}       

\section{Introduction}\label{sec1}
 The {\em eightfold way} of hadron spectra is the consequence of the
 SU(3) flavour symmetry in QCD which follows from the existence of
 three light quarks ($u$, $d$ and $s$).
 Numerical simulations of QCD have to take into account this basic fact.
 The ``quenched approximation'', which corresponds to the limit of
 infinitely heavy virtual quarks, is not reliable because of its
 unknown systematic errors (see, for instance,
 \cite{G-P:EPSILON,KPIPI}).

 Chiral perturbation theory (ChPT) \cite{CHPT} is very useful for
 extrapolating the results of numerical simulations to the small $u$-
 and $d$-quark masses but, of course, it has a finite range of
 applicability.
 In practice this means that one has to stay in a region where the
 one-loop ChPT formulas give a good approximation.
 For investigating the masses and couplings of pseudoscalar mesons one
 should reach the quark mass range below about one quarter of the
 strange quark mass ($m_{ud} \leq \frac{1}{4}m_s$)
 \cite{SHARPE-SHORESH,MPIMQ}.
 Present large scale QCD simulations -- especially with Wilson-type
 quark actions -- are not yet at small enough quark masses for the
 applicability of chiral perturbation theory.
 This is clearly displayed, for instance, by the absence of curvature
 (``no chiral logs'') in the present data about the dependence of
 $m_\pi^2$ and $f_\pi$ versus the quark mass $m_q$ (see, for instance,
 \cite{CPPACS:Alikhan,UKQCD:Allton,JLQCD:CHLOGM,JLQCD:CHLOGF}).

 Going to light quark masses in unquenched QCD simulations is a great
 challenge for computations because known algorithms have a
 substantial slowing down towards small quark masses.
 The present status has been summarized by the contributors to the panel
 discussion at the Berlin lattice conference \cite{Simulations:BERLIN}.
 In these contributions the computational costs were established at
 relatively large quark masses and extrapolated into the interesting
 range of small quark masses.
 Recent developments show, however, that at small quark masses
 new difficulties appear which might invalidate this extrapolation
 \cite{HMC:problems,CPPACS:Namekawa}.
 This was the motivation of our collaboration to investigate in a series
 of algorithmic studies the computational cost of numerical simulations
 down to about one fifth of the strange quark mass
 ($m_{ud} \leq \frac{1}{5}m_s$) \cite{NF2TEST,NF2BOSTON}.
 In our tests we used the two-step multi-boson algorithm \cite{TSMB}.

\section{Cost estimates}\label{sec2}
 The computational cost of obtaining a new, independent gauge
 configuration in an updating sequence with dynamical quarks can be
 parametrized, for instance, as
\be\label{eq01}
C = F\; (r_0 m_\pi)^{-z_\pi} \left(\frac{L}{a}\right)^{z_L}
\left(\frac{r_0}{a}\right)^{z_a} \ .
\ee
 Here $r_0$ is the Sommer scale parameter, $m_\pi$ the pion mass, $L$
 the lattice extension and $a$ the lattice spacing.
 The powers $z_{\pi,L,a}$ and the overall constant $F$ are empirically
 determined.
\begin{figure}[ht]
\vspace*{-10mm}
\begin{center}
\epsfig{file=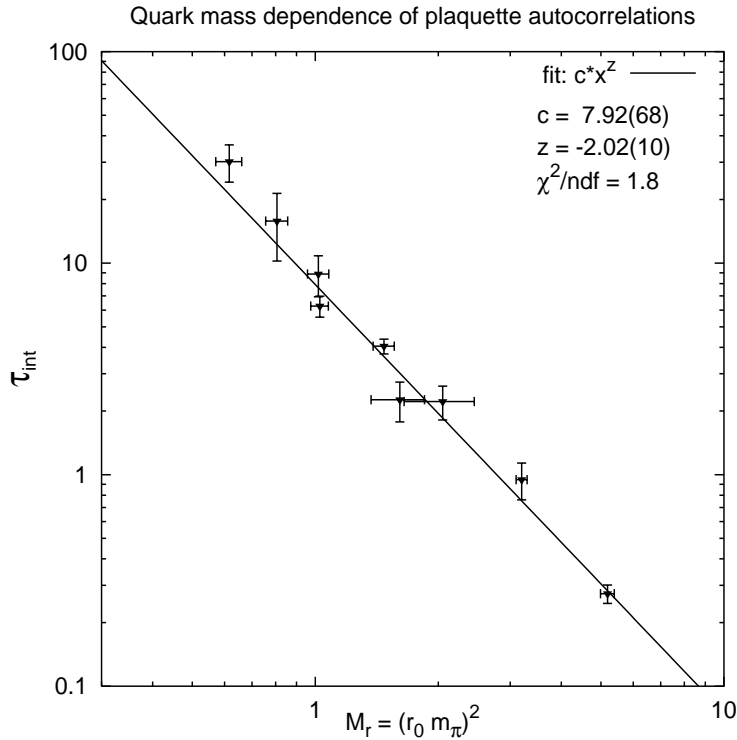,
        width=100mm,height=140mm,angle=-90,
        bbllx=50pt,bblly=-20pt,bburx=554pt,bbury=700pt}
\vspace*{-2mm}
\caption{\label{fig01}\em
 Power fit of the plaquette autocorrelation given in units of
 $10^6 \cdot {\rm MVM}$ as a function of the dimensionless quark mass
 parameter $M_r$.
 The best fit of the form $c M_r^z$ is at $c=7.92(68),\; z=-2.02(10)$.}
\end{center}
\vspace*{-5mm}
\end{figure}
\begin{figure}[ht]
\vspace*{-10mm}
\begin{center}
\epsfig{file=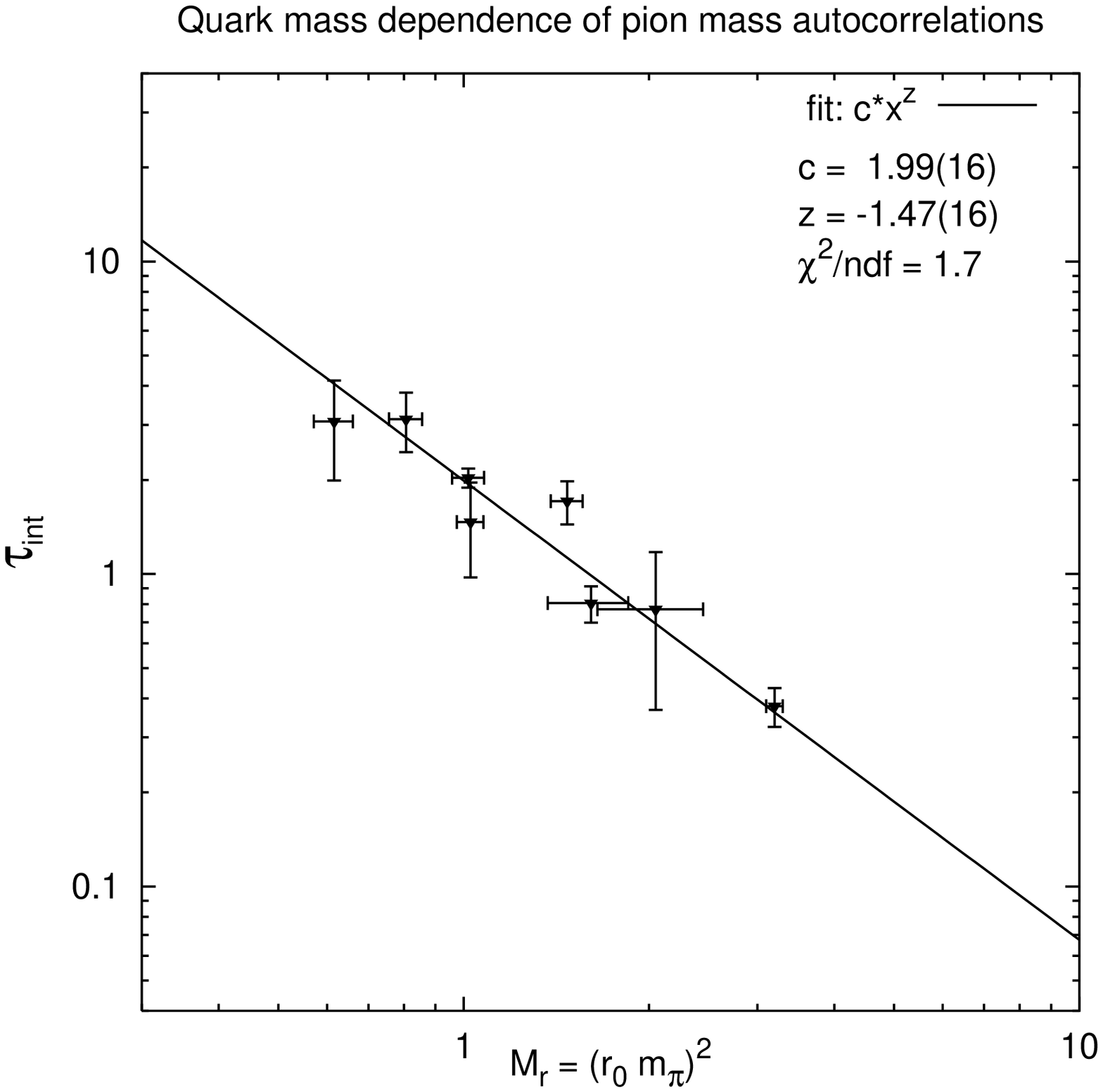,
        width=100mm,height=140mm,angle=-90,
        bbllx=50pt,bblly=-20pt,bburx=554pt,bbury=700pt}
\vspace*{-2mm}
\caption{\label{fig02}\em
 Power fit of the autocorrelation of the pion mass given in units of
 $10^6 \cdot {\rm MVM}$ as a function of the dimensionless quark mass
 parameter $M_r$.
 The best fit of the form $c M_r^z$ is at $c=1.99(16),\; z=-1.47(16)$.}
\end{center}
\vspace*{-5mm}
\end{figure}

 In order to limit the necessary computer time our collaboration first
 performed high statistics simulations with $N_f=2$ degenerate quark
 flavours on coarse lattices ($8^3 \cdot 16$) with a lattice spacing
 $a\simeq 0.27\, {\rm fm}$.
 ($a$ is determined by $r_0 \simeq 0.5\, {\rm fm}$ from the mesured
 value of $r_0/a$.)
 This gives for the space-extension $L \simeq 2.2\, {\rm fm}$, which is
 sufficient for keeping finite volume effects tolerable, for instance,
 for $m_\pi$ and $f_\pi$.
 For the definition of the quark mass the dimensionless quantity
\begin{equation}\label{eq02}
M_r \equiv (r_0 m_\pi)^2
\end{equation}
 has been used, which already appears in (\ref{eq01}).
 In this variable the strange quark mass can be defined as
 $M_{r,strange} \equiv 3.1$.

 Our results for the integrated autocorrelations of different physical
 quantities ($\tau_{int}$) have been described in detail in
 \cite{NF2TEST}.
 For illustration see figures \ref{fig01} and \ref{fig02} where
 the quark mass dependence of $\tau_{int}^{plaquette}$ and
 $\tau_{int}^{m_\pi}$ are shown, together with power fits.
 The units for the costs are $10^6$
 fermion-matrix-vector-multiplications (MVMs).
 Comparing the two figures one sees that the ``cost'' for $m_\pi$
 is substantially less than the one for the average plaquette.
 The fitted powers of the quark mass are also different.
 In the notation of the formula in (\ref{eq01}) $\tau_{int}^{plaquette}$
 gives $z_\pi \simeq 4$, whereas $\tau_{int}^{m_\pi}$ is consistent with
 $z_\pi \simeq 3$.
 The pion coupling constant $f_\pi$, which is related to the constant in
 front of the pion contribution in the pseudoscalar propagator, has
 even shorter autocorrelations than $m_\pi$.
 This is obviously rather advantageous for obtaining physical
 information from the quark mass dependence of the pion mass and
 coupling (see \cite{SHARPE-SHORESH}).
\begin{table*}
\caption{\em Runs for comparing the simulations costs at different
 volumes.
 For the definition of algorithmic parameters see
 \protect{\cite{NF2TEST}}.
\label{tab01}}
\begin{center}
\begin{tabular}{|c|c|c|c|c|c|c|c|c|}
\hline
run & lattice & $\beta$ & $\kappa$ & 
$n_1$ & $n_2$ & $n_3$  & $\lambda$ & $\epsilon\cdot 10^4$
\\ \hline
(e) & $8^3 \cdot 16$    & 4.76 & 0.190 & 44 & 360 & 380  & 3.6 & 2.7
\\ \hline
(e16) & $16^4$          & 4.76 & 0.190 & 72 & 350 & 440  & 3.6 & 2.7
\\ \hline
(f) & $8^3 \cdot 16$    & 4.80 & 0.190 & 44 & 360 & 380  & 3.6 & 2.7
\\ \hline
(f12) & $12^3 \cdot 24$ & 4.80 & 0.190 & 72 & 500 & 560  & 3.4 & 1.36
\\ \hline
(h) & $8^3 \cdot 16$    & 4.68 & 0.195 & 66 & 900 & 1200 & 3.6 & 0.36
\\ \hline
(h16) & $16^4$          & 4.68 & 0.195 & 96 & 860 & 1100 & 3.6 & 0.36
\\ \hline
\end{tabular}
\end{center}
\end{table*}

 Our collaboration also studied the volume dependence of the cost by
 choosing a few $8^3 \cdot 16$ runs and repeating them on either
 $16^4$ or $12^3 \cdot 24$ lattices, without changing other parameters
 \cite{NF2BOSTON}.
 The parameters of the chosen points are collected in table \ref{tab01}
 and some results are shown in table \ref{tab02}.
 The error analysis and integrated autocorrelations in the runs on the
 larger lattices have been obtained using the {\em linearization method}
 of the ALPHA collaboration \cite{ALPHA:BENCHMARK}.
 The conclusion from these tests is that the cost increase with the
 lattice volume is quite acceptable because it is close to the trivial
 volume factor.
 In case of the autocorrelation of the pion mass the observed increase
 turns out to be even smaller.
 This is partly due to the intrinsic fluctuation present in the
 pion propagator which is originated from the freedom of randomly
 choosing the position of the source.

\section{Chiral logs?}\label{sec3}
 The behaviour of physical quantities, as for instance the pseudoscalar
 meson (``pion'') mass $m_\pi$ or pseudoscalar decay constant $f_\pi$
 as a function of the quark mass are characterized by the appearance of
 {\em chiral logarithms}.
 These chiral logs, which are due to virtual pseudoscalar meson loops,
 have a non-analytic behaviour near zero quark mass of a generic form
 $m_q\log m_q$.
 They imply relatively fast changes of certain quantities near zero
 quark mass which are not seen in present data \cite{CPPACS:Alikhan,
 UKQCD:Allton,JLQCD:CHLOGM,JLQCD:CHLOGF,CPPACS:Namekawa}.
\begin{table*}
\caption{\em Results from the runs with parameters shown in table
 \protect{\ref{tab01}}.
 The lattice extension $L$ is obtained from the value of $r_0/a$.
 $M_r$ is the quark mass parameter defined in \protect{(\ref{eq01})}.
 The cost of an update cycle $C_{uc}$ and the integrated
 autocorrelations of the average plaquette and of the pion mass
 are given, respectively, in $10^3$ MVMs and flops.
\label{tab02}}
\begin{center}
\begin{tabular}{|c|c|c|c|r|r|}
\hline
run & $L\,[{\rm fm}]$ & $M_r$ & $C_{uc}\,[{\rm kMVM}]$ & 
\multicolumn{1}{|c|}{$\tau_{int}^{plaq}\,[{\rm flop}]$} & 
\multicolumn{1}{|c|}{$\tau_{int}^{m_\pi}\,[{\rm flop}]$}
\\ \hline
(e)   & 2.31(6) & 1.473(88) & 8.50 & 4.59(37) $\cdot 10^{13}$ & 
1.94(31) $\cdot 10^{13}$
\\ \hline
(e16) & 4.57(9) & 1.401(79) & 12.4 & 7.5(1.3) $\cdot 10^{14}$ & 
5.02(55) $\cdot 10^{13}$
\\ \hline
(f)   & 2.25(4) & 1.026(51) & 8.48 & 7.47(84) $\cdot 10^{13}$ & 
1.76(59) $\cdot 10^{13}$
\\ \hline
(f12) & 3.02(9) & 1.37(9)   & 12.3 & 2.40(41) $\cdot 10^{14}$ & 
4.52(82) $\cdot 10^{13}$
\\ \hline
(h)   & 2.27(5) & 0.806(50) & 16.2 & 1.7(6) $\cdot 10^{14}$   & 
3.3(7) $\cdot 10^{13}$
\\ \hline
(h16) & 4.1(3)  & 0.93(19)  & 23.7 & 1.10(17) $\cdot 10^{15}$ & 
8.3(8) $\cdot 10^{13}$
\\ \hline
\end{tabular}
\end{center}
\end{table*}

 In our range of quark masses ($m_q \simeq \frac{1}{5}m_s$) the
 logarithmic curvature due to chiral logs has to appear, at least in
 the continuum limit.
 Although we have rather coarse lattices ($a \simeq 0.27\, {\rm fm}$)
 and, in addition, up to now we are working with unrenormalized
 quantities -- without the $Z$-factors of multiplicative renormalization
 -- it is interesting to see whether the effects of chiral logs can
 already be seen in our data.
\begin{figure}[th]
\begin{center}
\epsfig{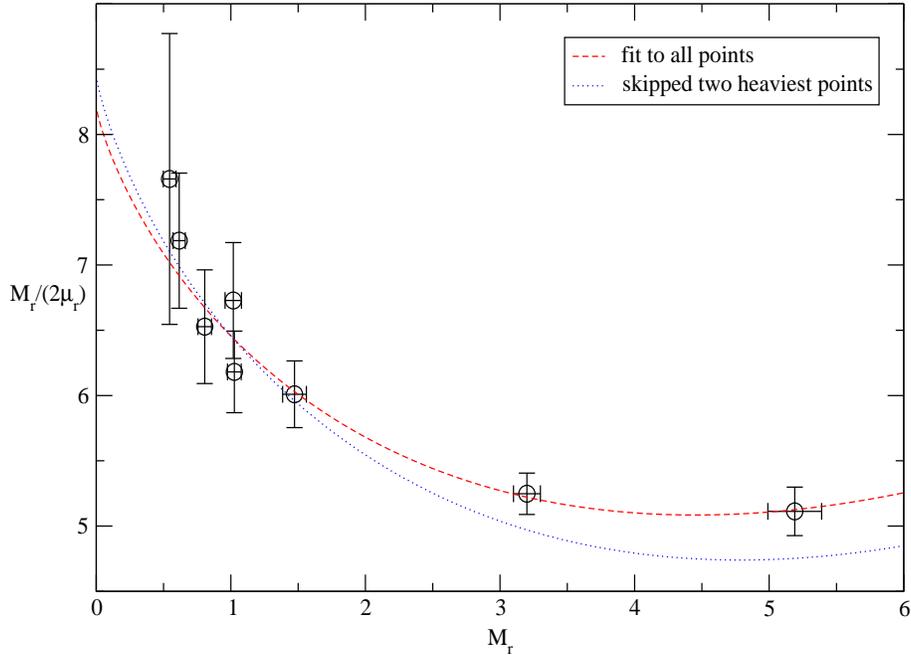}
\vspace*{-2mm}
\caption{\label{fig03}\em
 Fits of the pseudoscalar meson mass-squared with the one-loop ChPT
 formula.}
\end{center}
\end{figure}
\begin{figure}[th]
\begin{center}
\epsfig{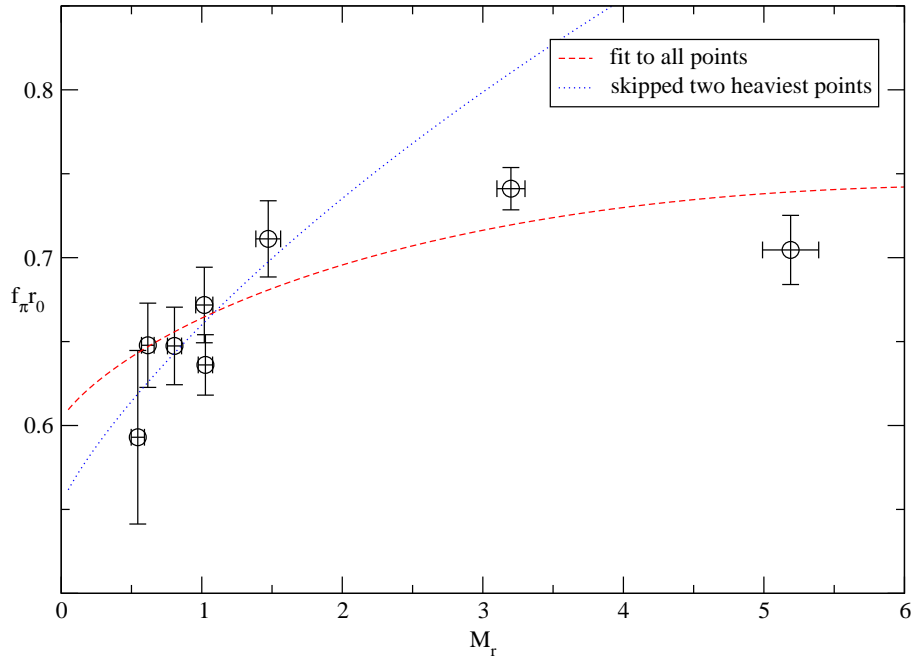}
\vspace*{-2mm}
\caption{\label{fig04}\em
 Fits of the pseudoscalar meson decay constant with the one-loop ChPT
 formula.}
\end{center}
\end{figure}

 Let us recall the one-loop ChPT formulas for $m_\pi^2$:
\begin{equation}\label{eq03}
\frac{M_r}{2\mu_r} = Br_0 - \frac{M_rBr_0}{16\pi^2 (f r_0)^2}
\log\frac{(\Lambda_3r_0)^2}{M_r} + {\cal O}(M_r^2) \ ,
\end{equation}
 and for $f_\pi$:
\begin{equation}\label{eq04}
f_\pi r_0 = fr_0 + \frac{M_r}{8\pi^2fr_0}
\log\frac{(\Lambda_4r_0)^2}{M_r} + {\cal O}(M_r^2) \ .
\end{equation}
 These are deduced from the formulas in ref.~\cite{LEUTWYLER,MPIMQ},
 with our convention $f_\pi^{physical} \simeq 131\, {\rm MeV}$.
 Besides $M_r$ defined in (\ref{eq01}), the quark mass defined by the
 PCAC relation $m_q^{PCAC}$ also appears in the dimensionless
 combination
\begin{equation}\label{eq05}
\mu_r \equiv m_q^{PCAC}r_0 \ .
\end{equation}
 (For the definition of $m_q^{PCAC}$ see, for instance, the equations
 (20)-(23) in our paper \cite{NF2TEST}.)
 The dimensionless parameters $Br_0$, $fr_0$ and $\Lambda_3 r_0$ in
 (\ref{eq03}) and $fr_0$ and $\Lambda_4 r_0$ in (\ref{eq04}) have to be
 fitted to the data.

 The data from table 3 of ref. \cite{NF2TEST} and the one-loop ChPT
 fits are shown in figures \ref{fig03} and \ref{fig04}.
 (The runs with label (c) and (d), which have low statistics and hence
 large statistical errors, are omitted here.)
 In both figures two fits are shown: one taking into account all points
 and another one where the points above $M_r=2$ are omitted.
 As the figures show, the data in the small quark mass range clearly
 show the expected qualitative behaviour with chiral logs.
 The fit parameters have reasonable values, similar to the ones
 deduced in \cite{MPIMQ} from previous lattice data at larger quark
 masses \cite{CPPACS:Alikhan,UKQCD:Allton}.
 For instance, the fits with all points correspond to the parameters:
 $Br_0 = 8.2$, $f r_0 = 0.27$, $\Lambda_3 r_0 = 3.5$ in formula
 (\ref{eq03}) and $f r_0 = 0.60$, $\Lambda_4 r_0 = 4.3$ in formula
 (\ref{eq04}).

 The fact that our data in the small quark mass range
 $\frac{1}{5}m_s \leq m_q \leq \frac{2}{3}m_s$ show the expected
 logarithmic behaviour of chiral perturbation theory is quite
 satisfactory.
 However, since we are far from the continuum limit and we did not yet
 take into account the $Z$-factors of multiplicative renormalization,
 further work is needed to finally deduce the physical values of the
 fit parameters.

\section{Eigenvalue spectra}\label{sec4}
 The eigenvalue spectrum of the Wilson-Dirac matrix is interesting
 both physically and from the point of view of simulation algorithms.
 From the algorithmic point of view the knowledge of low-lying
 eigenvalues is crucial for the optimization of polynomial
 approximations in TSMB.
 An interesting question is whether there is a statistically significant
 presence of configurations with negative determinant, i.~e. with an
 odd number of negative eigenvalues of the Wilson-Dirac matrix.
 A significant number of configurations with negative determinant would
 be a serious obstacle for performing numerical simulations with and
 odd number of flavours -- as it occurs with $u$-, $d$- and $s$-quarks
 in nature.
 The results of our collaboration show \cite{NF2TEST,NF2BOSTON} that
 in the investigated range of quark masses no {\em sign problem}
 occurs because the statistical weight of configurations with negative
 determinant is negligible even at our smallest quark masses.

\section{Conclusion and outlook}\label{sec5}
 Our cost estimates for numerical simulations of QCD with light
 dynamical Wilson-quarks -- as a function of the quark mass and volume
 -- allow a reasonably precise prediction of the computational costs
 for future simulations.
 Up to now we did not yet perform a systematic investigation of the
 slowing down with decreasing lattice spacing, expressed by the power
 $z_a$ in (\ref{eq01}), but previous experience in the supersymmetric
 Yang-Mills theory \cite{SYMREV,SYMMUNST} at small lattice spacings
 $a \simeq 0.06\, {\rm fm}$ and our preliminary results in QCD near
 $a \simeq 0.17\, {\rm fm}$ are consistent with the expected behaviour
 $z_a = 2$ \cite{Simulations:BERLIN}.

 Assuming $z_a = 2$ and starting from the results of runs $(h)$ and
 $(h16)$ in table \ref{tab02} one can deduce the following estimates
 for obtaining 100 independent gauge configurations:
\begin{itemize}
\item
 On a $24^3 \cdot 48$ lattice with lattice spacing $a = 0.125\,{\rm fm}$
 at quark mass parameter $M_r = 0.8$ corresponding to
 $m_{ud} \simeq \frac{1}{4}m_s$ and physical extension characterized by
 $Lm_\pi \simeq 6$ the cost would be
 $C \simeq 9\cdot 10^{16}-3\cdot 10^{18}\, {\rm flop}$, depending on
 whether one considers $\tau_{int}^{m_\pi}$ or $\tau_{int}^{plaq}$ as
 relevant.
\item
 Similarly, on a $32^3 \cdot 64$ lattice with lattice spacing
 $a = 0.06\,{\rm fm}$ at the same quark mass and $Lm_\pi \simeq 4$
 the cost would be
 $C \simeq 6\cdot 10^{17}-4\cdot 10^{19}\, {\rm flop}$.
\end{itemize}
 Compared to most of the estimates in \cite{Simulations:BERLIN} these
 numbers are lower.
 Note that the costs for parameter tuning and equilibration --
 which are not negligible -- are not included in these estimates.
 Nevertheless, the necessary order of magnitude of CPU times for these
 kinds of calculations is expected to be available for the lattice
 community in forthcoming years.




\begin{thebibliography}{99}
%
\bibitem{G-P:EPSILON}
M. Golterman and E. Pallante,
\newblock hep-lat/0208069.
%
\bibitem{KPIPI}
C.J.D. Lin, G. Martinelli, E. Pallante, C.T. Sachrajda and G. Villadoro,
\newblock hep-lat/0209107.
%
\bibitem{CHPT}
J. Gasser and H. Leutwyler,
\newblock Annals Phys. {\bf 158} (1984) 142.
%
\bibitem{SHARPE-SHORESH}
S.R. Sharpe and N. Shoresh,
\newblock Phys. Rev. {\bf D62} (2000) 094503; hep-lat/0006017.
%
\bibitem{MPIMQ}
S. D\"urr,
\newblock hep-lat/0208051. 
%
\bibitem{CPPACS:Alikhan}
CP-PACS Collaboration, A. Ali-Khan et al.,
\newblock Phys. Rev. {\bf D65} (2002) 054505; hep-lat/0105015.
%
\bibitem{UKQCD:Allton}
UKQCD Collaboration, C.R. Allton et al.,
\newblock Phys. Rev. {\bf D65} (2002) 054502; hep-lat/0107021.
%
\bibitem{JLQCD:CHLOGM}
JLQCD Collaboration, S. Aoki et al.,
\newblock Nucl. Phys. Proc. Suppl. {\bf 106} (2002) 224; 
hep-lat/0110179.
%
\bibitem{JLQCD:CHLOGF}
JLQCD Collaboration, S. Hashimoto et al.,
\newblock hep-lat/0209091.
%
\bibitem{Simulations:BERLIN}
Contributions of N.H. Christ, S. Gottlieb, K. Jansen, Th. Lippert,
A. Ukawa and H. Wittig,
\newblock Nucl. Phys. Proc. Suppl. {\bf 106} (2002).
%
\bibitem{HMC:problems}
UKQCD Collaboration, B. Joo et al.,
\newblock Nucl. Phys. Proc. Suppl. {\bf 106} (2002) 1073;
hep-lat/0110047.
%
\bibitem{CPPACS:Namekawa}
CP-PACS Collaboration, Y. Namekawa et al.,
\newblock hep-lat/0209073.
%
\bibitem{NF2TEST}
qq+q Collaboration, F. Farchioni, C. Gebert, I. Montvay and L. Scorzato,
\newblock hep-lat/0206008.
%
\bibitem{NF2BOSTON}
qq+q Collaboration, F. Farchioni, C. Gebert, I. Montvay and L. Scorzato,
\newblock hep-lat/0209038.
%
\bibitem{TSMB}
I. Montvay,
\newblock Nucl. Phys. {\bf B466} (1996) 259; hep-lat/9510042.
%
\bibitem{ALPHA:BENCHMARK}
ALPHA Collaboration, R. Frezzotti, M. Hasenbusch, U. Wolff, J. Heitger
and K. Jansen,
\newblock Comput. Phys. Commun. {\bf 136} (2001) 1; hep-lat/0009027.
%
\bibitem{LEUTWYLER}
H. Leutwyler,
\newblock Nucl. Phys. Proc. Suppl. {\bf 94} (2001) 108; hep-ph/0011049.
%
\bibitem{SYMREV}
I. Montvay, 
\newblock Int. J. Mod. Phys. A17 (2002) 2377; hep-lat/0112007.
%
\bibitem{SYMMUNST}
R. Peetz, F. Farchioni, C. Gebert, G. M\"unster,
\newblock hep-lat/0209065.
%
\end{thebibliography}
\end{document}